\documentclass{emulateapj}
\usepackage{apjfonts}

\newcommand{\etal}{et~al.}
\newcommand{\MgIIdblt}{{\rm Mg}\kern 0.1em{\sc ii}~$\lambda\lambda 2796, 2803$}
\newcommand{\MgII}{\hbox{{\rm Mg}\kern 0.1em{\sc ii}}}
\newcommand{\HI}{\hbox{{\rm H}\kern 0.1em{\sc i}}}

\newcommand{\cmsq}{\hbox{cm$^{-2}$}}

\newcommand{\fdmax}{\hbox{$f_c ( \leq\!\! D )$}}
\newcommand{\fdrmax}{\hbox{$f_c (\!\leq\!\! D/R_{\rm vir})$}}
\newcommand{\fdbar}{\hbox{$f_c ( D\, )$}}
\newcommand{\fdrbar}{\hbox{$f_c ( D/R_{\rm vir} )$}}

\newcommand{\magiicat}{\hbox{{\rm MAG}{\sc ii}CAT}}

\shorttitle{\sc {\MgII}-Absorber Halo Masses}
\shortauthors{\sc Churchill et~al.}
\slugcomment{Submitted to ApJL, November 2, 2012}

\begin{document}


\title{The Self-Similarity of the Circumgalactic 
Medium with \\ Galaxy Virial Mass: Implications for Cold-Mode Accretion}


\author{\sc
Christopher W. Churchill\altaffilmark{1},
Nikole M. Nielsen\altaffilmark{1},
Glenn G. Kacprzak\altaffilmark{2,3},
and
Sebastian Trujillo-Gomez\altaffilmark{1}
}
\altaffiltext{1}{New Mexico State University, Las Cruces, NM 88003}
\altaffiltext{2}{Swinburne University of Technology, Victoria 3122, Australia}
\altaffiltext{3}{Australian Research Council Super Science Fellow}

\begin{abstract}

We apply halo abundance matching to obtain galaxy virial masses,
$M_{\rm\,h}$, and radii, $R_{\rm vir}$, for 183 ``isolated''
galaxies from the ``{\MgII} Absorber-Galaxy Catalog'' ({\magiicat}).
All galaxies have spectroscopic redshifts ($0.07 \leq z \leq 1.12$)
and their circumgalactic medium (CGM) is probed in {\MgII} absorption
within projected galactocentric distances $D\leq 200$~kpc. We examine
the behavior of equivalent width, $W_r(2796)$, and covering fraction,
$f_c$, as a function of $D$, $D/R_{\rm vir}$, and $M_{\rm\,h}$.
Bifurcating the sample at the median mass $\log M_{\rm\,h}/M_{\odot} =
12$, we find: [1] systematic segregation of $M_{\rm\,h}$ on the
$W_r(2796)$--$D$ plane ($4.0~\sigma$); high-mass halos are found at
higher $D$ with larger $W_r(2796)$ compared to low-mass halos.  On the
$W_r(2796)$--$D/R_{\rm vir}$ plane, mass segregation vanishes and we
find $W_r(2796) \propto (D/R_{\rm vir})^{-2}$ ($8.9~\sigma$); [2]
high-mass halos have larger $f_c$ at a given $D$, whereas $f_c$ is
independent of $M_{\rm\,h}$ at all $D/R_{\rm vir}$; [3] $f_c$ is
constant with $M_{\rm\,h}$ over the range $10.7 \leq \log M_{\rm
\,h}/M_{\odot} \leq 13.9$ within a given $D$ or $D/R_{\rm vir}$.  The
combined results suggest the {\MgII} absorbing CGM is self-similar
with halo mass, even above $\log M_{\rm\,h}/M_{\odot} \simeq 12$,
where cold mode accretion is predicted to be quenched.  If theory is
correct, either outflows or sub-halos must contribute to absorption in
high-mass halos such that low- and high-mass halos are observationally
indistinguishable using {\MgII} absorption strength once impact
parameter is scaled by halo mass.  Alternatively, the data may
indicate predictions of a universal shut down of cold-mode accretion
in high-mass halos may require revision.
\end{abstract}


\section{Introduction}
\label{sec:intro}

Direct observation of the circumgalactic medium (CGM) is important for
exposing the link between the intergalactic medium and galaxies, and
the processes governing their star formation histories, stellar
masses, luminosities, and morphologies.  The CGM harbors a reservoir
of chemically enriched gas with a mass that may rival the gas
reservoir in galaxies \citep{tumlinson11}.  The {\MgIIdblt} absorption
doublet observed in quasar spectra is an ideal probe of the CGM
\citep[see][for a review]{cwc-china}; it traces low-ionization gas
over the broad range $10^{16.5} \leq N(\HI) \leq 10^{21.5}$~{\cmsq}
\citep{weakI, archiveI, rao00, weakII} and is detected out to
projected distances of $\sim 150$ proper kpc \citep{kacprzak08,
chen10a, churchill12, nielsen12}.

In general, the CGM is a complex dynamical region affected by
accretion, winds, and mergers.  Two distinct modes of accretion are
theorized to operate, ``hot'' or ``cold'', where the mode depends on
whether the halo mass, $M_{\rm\,h}$, is above or below a critical
threshold $M_{\rm crit}$, where $\log M_{\rm crit}/M_{\odot} \sim 12$
\citep[e.g.,][]{birnboim03, keres05, dekel06, keres09, stewart11,
vandevoort11}.  Cold-mode accretion is predicted in $M_{\rm\,h} <
M_{\rm crit}$ halos, whereas hot-mode is predicted in $M_{\rm\,h} >
M_{\rm crit}$ halos.  Though some observations provide plausible
evidence for cold-mode accretion at $z<1$ \citep{kcems, ribaudo11,
thom11, ggk-q1317}, the baryonic mass in the cold mode is expected to
diminish with decreasing redshift.

If accretion dominates, the distribution of absorber equivalent
widths, $W_r(2796)$, and the absorption covering fraction, $f_c$, are
predicted to markedly decline in $M_{\rm\,h} > M_{\rm crit}$ halos.
If a large reservoir of cold gas ($T \sim 10^4$--$10^5$~K) is present
in the CGM of $M_{\rm\,h}> M_{\rm crit}$ halos, it could imply
outflows \citep[cf.,][]{stewart11}.  Indeed, winds are commonly
observed in {\MgII} absorption \citep[e.g.,][]{tremonti07, martin09,
weiner09, rubin10, martin12}.  Alternatively, the increased number of
sub-halos associated with higher mass halos \citep{klypin11} could
counteract the predicted behavior of $W_r(2796)$ and $f_c$.

To observationally examine the validity of the hot/cold accretion
theoretical paradigm, we require the halo masses of galaxies
associated with absorption systems.  A robust method for determining
these masses is the parameterless halo abundance matching technique
\citep[e.g.,][]{conroy06, conroy09, behroozi10, trujillo-gomez11,
rodriguez-puebla12}.  The method has been extremely successful at
reproducing the two-point correlation function with redshift,
luminosity, and stellar mass \citep{conroy06, trujillo-gomez11}, the
galaxy velocity function, and the luminosity-velocity and baryonic
Tully-Fisher relations \citep{trujillo-gomez11}.

In this {\it Letter}, we explore the connection between $M_{\rm\,h}$
and the {\MgII} absorbing CGM and show that the ``cold'' CGM is
self-similar with halo mass over the mass range $10.7 \leq \log M_{\rm
\,h}/M_{\odot} \leq 13.9$.  For this work, we define halo mass as the
galaxy virial mass, including dark matter and baryons.  In
\S~\ref{sec:HAM} we describe our galaxy sample and our method to
estimate each galaxy's halo mass.  We present our findings in
\S~\ref{sec:results}.  In \S~\ref{sec:conclude}, we summarize and
provide concluding remarks. Throughout, we adopt a $h=0.70$,
$\Omega_{\rm M}=0.3$, $\Omega_{\Lambda}=0.7$ flat cosmology.

\begin{figure*}[thb]
\epsscale{1.1}  
\plotone{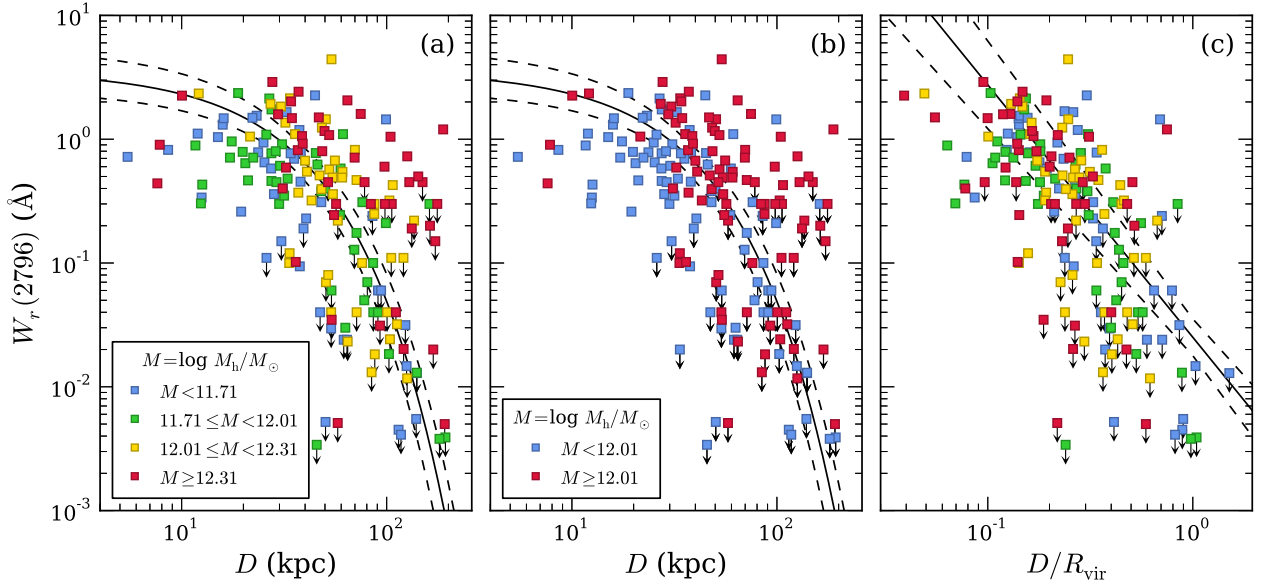}
\caption{$W_r(2796)$ versus $D$ and $D/R_{\rm vir}$.  Limits are
indicated with downward arrows. The solid curve is the fit to the full
sample and dashed curves are the $1~\sigma$ envelope.  (a) The full
sample of galaxies separated into $M_{\rm\,h}$ quartiles.  (b) The
high-mass and low-mass subsamples.  The high- and low-mass subsamples
with measured absorption are segregated on this diagram with
$4.0~\sigma$ significance. (c) $W_r(2796)$ versus $D/R_{\rm vir}$,
which is anti-correlated with $8.9~\sigma$ significance.  The fit
yields $W_r(2796) \propto (D/R_{\rm vir})^{-2}$.  The mass segregation
vanishes.}
\label{fig:EWvsD}
\end{figure*}

\section{The Sample, Viral Masses, and Virial Radii}
\label{sec:HAM}

Our sample comprises 183 ``isolated'' galaxies from the ``{\MgII}
Absorber-Galaxy Catalog'' \citep[{\magiicat},][]{nielsen13}.  Each
galaxy has a spectroscopic redshift ($0.07 \leq z \leq 1.12$).  The
impact parameter range is $5.4 \leq D \leq 194$~kpc.  The ranges of
the AB absolute $B$- and $K$- band magnitudes are $-16.1 \geq M_B \geq
-23.1$ and $-16.9 \geq M_K \geq -25.3$, with rest-frame $B\!-\!K$
colors $0.04 \leq B\!-\!K \leq 4.09$.  Including $3~\sigma$ upper
limits, the rest-frame {\MgII} $\lambda 2796$ equivalent widths have
the range $0.0034 \leq W_r(2796) \leq 2.90$~{\AA} with one system at
$W_r(2796) = 4.42$~{\AA}.


The $M_{\rm\,h}$ for each galaxy was obtained by abundance matching
halos in the Bolshoi $N$-body cosmological simulation \citep{klypin11}
to the observed $r$-band luminosity function (LF) from the COMBO-17
survey \citep{wolf03}.  In short, the method links the luminosity of
galaxies to halo properties in a monotonic fashion, reproducing the LF
by construction.  Following \citet{trujillo-gomez11}, we adopt the
maximum circular velocity, $V_c^{\rm max}$, and solve for the unique
relation $n(\!<\!M_r) = n(\!<\!V_c^{\rm max})$, which properly
accounts for the depth of the potential well and is unambiguously
defined for both central halos and sub-halos.

The halos were matched in the five redshift bins centered at $z=0.3$,
$0.5$, $0.7$, $0.9$, $1.1$ of the COMBO-17 $r$-band LF.  The galaxy
$r$-band absolute magnitudes, $M_r$, were determined by $k$-correcting
\citep[e.g.,][]{kim96} the $R$, SDSS $r$, or {\it HST}/WFPC
F702W observed magnitudes using the actual filter response curves.  We
employed \citet{cww80} spectral energy distribution (SED) templates
from \citet{bolzonella00}.  The adopted SED for each galaxy was
obtained by matching its observed color to the redshifted SEDs.  The
resulting range of $M_r$ is $-15.1 \geq M_r \geq -22.7$.

Scatter between $M_{\rm\,h}$ and $M_r$ originates from scatter in the
$M_{\rm\,h}$--$V_c^{\rm max}$ relation due to different formation
times of halos with similar mass.  After mapping $M_{\rm\,h}$ to
$M_r$, we account for this scatter by computing the average $M_{\rm
\,h}$ in a $\Delta M_{\rm r} = 0.1$ bin centered on the $M_r$ of the
galaxy. Thus, each derived $M_{\rm\,h}$ is interpreted as the average
halo mass of a galaxy with $M_r$.  Varying the bin size $\Delta M_r$
had virtually no systematic effect nor change in the scatter of each
mass estimate.

The primary uncertainty in the derived $M_{\rm\,h}$ is the observed
LF.  Tests bracketing reasonable extremes of possible systematic
errors in the measured LF yield $M_{\rm\,h}$ that are qualitatively
unchanged.  Full details of our methods and uncertainties are
described in \citet{churchill13}.  The masses have the range $10.74
\leq \log M_{\rm\,h}/M_{\odot} \leq 13.89$ with median $\log
M_{\rm\,h}/M_{\odot} = 12.01$. Including systematics and scatter,
the uncertainties are $\delta \log M_{\rm\,h} \simeq 0.1$ at $\log
M_{\rm\,h}/M_{\odot}=10$ increasing quasi-linearly to $\delta \log
M_{\rm\,h} \simeq 0.35$ at $\log M_{\rm\,h}/M_{\odot}=13$.

We obtain the virial radius, $R_{\rm vir}$, for each galaxy using the
formulae of \citet{bryan98},
\begin{equation}
R^3_{\rm vir} = 
\frac{3}{4\pi} \frac{M_{\rm\,h}}{\rho_c \Delta _c(z)} 
\qquad 
\Delta _c(z) = \frac{(18\pi ^2 + 82x - 39x^2)}{1+x}
\label{eq:Rvir}
\end{equation}
where $x = \Omega _m(1+z)^3/[\Omega_m (1+z)^3 + \Omega _\Lambda ]$.
The resulting radii have the range $70 \leq R_{\rm vir} \leq 840$
proper kpc with uncertainty of $\delta R_{\rm vir}/R_{\rm vir} \simeq
0.1$, where the uncertainty in each $R_{\rm vir}$ accounts for the
standard deviation in the average virial mass assigned to each galaxy.

\begin{figure*}[thb]
\epsscale{1.1} 
\plotone{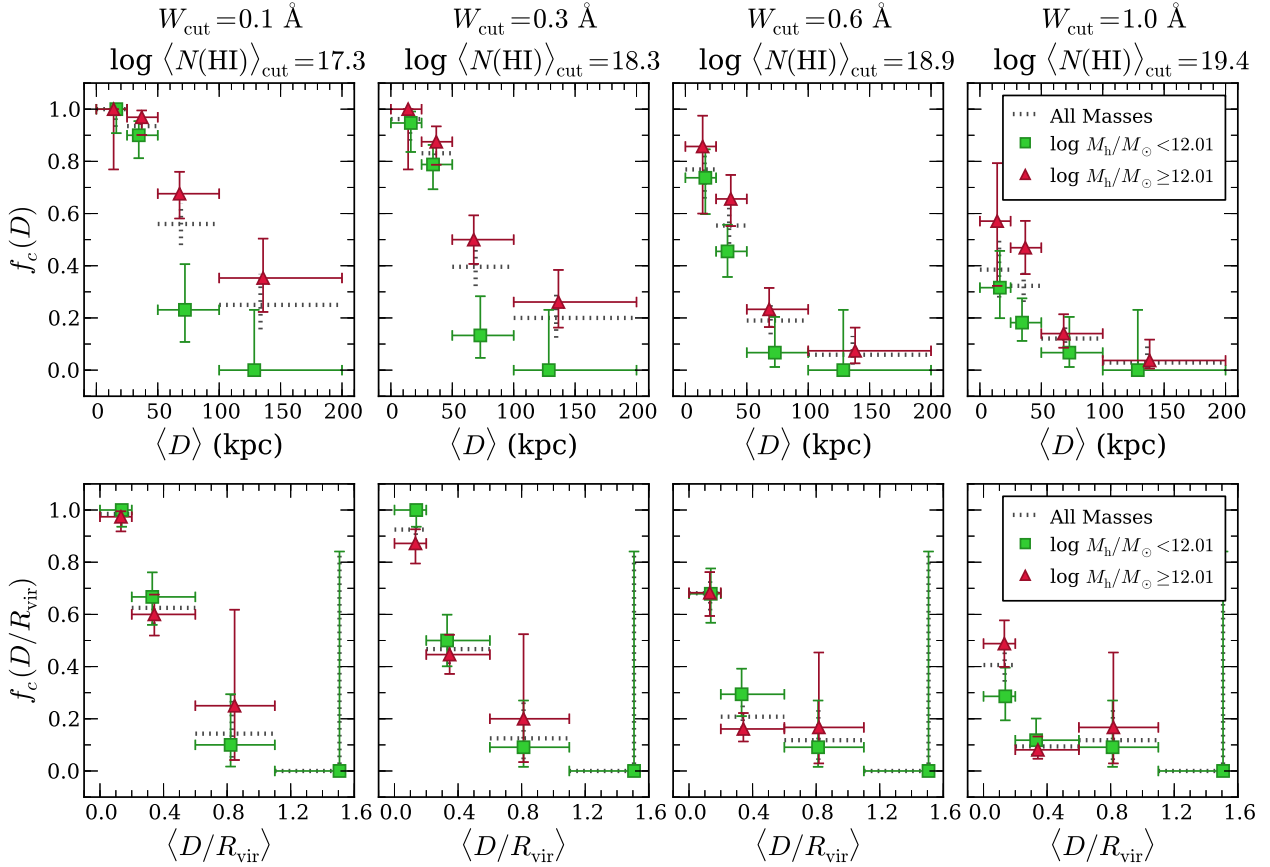}
\caption{(upper panels) The CGM covering fraction, {\fdbar}, for
$W_r(2796) \geq W_{\rm cut}$ in the $D$ bins (0,25], (25,50],
(50,100], and (100,200) kpc for the high- and low-mass subsamples
(dashed bars are the full sample). Each $W_{\rm cut}$ corresponds to a
mean minimum $\log N({\HI})$.  (lower panels) The covering fraction,
{\fdrbar}, for $W_r(2796) \geq W_{\rm cut}$ in the $D/R_{\rm vir}$
bins (0,0.2], (0.2,0.6], (0.6,1.1], and (1.1,1.5].  High-mass halos
tend to have larger {\fdbar}, especially for $D>50$ kpc and $W_{\rm
cut} =0.1$ and 0.3~{\AA}, whereas {\fdrbar} is independent of
$M_{\rm\,h}$, suggesting the {\MgII} absorbing CGM is self-similar
with halo mass.}
\label{fig:fDbar}
\end{figure*}

\section{Results}
\label{sec:results}


\subsection{$W_r(2796)$ versus $D$ and $D/R_{\rm vir}$}

The observed anti-correlation between $W_r(2796)$ and $D$ is firmly
established \citep[e.g.,][and references therein]{nielsen12}.
However, there is significant scatter in the relation.  The source of
the scatter has been investigated assuming a ``second parameter'',
i.e., galaxy luminosity \citep{kacprzak08, chen10a}, stellar mass or
specific star formation rate \citep{chen10b}, morphology
\citep{ggk-morphology}, or geometry and orientation
\citep{bouche11,kcems}.

We divided the sample into halo mass quartiles.  In
Figure~\ref{fig:EWvsD}a, we plot $W_r(2796)$ versus $D$ with points
colored by mass (see legend).  The curve is a log-linear fit
\citep{nielsen12}.  Generally, the highest-mass halos (yellow and red
points) segregate above the fitted curve, whereas the lowest-mass
halos (blue and green points) segregate below the fitted curve.

In Figure~\ref{fig:EWvsD}b, we plot the data split by the median
mass: the high-mass subsample (red points) has $\log
M_{\rm\,h}/M_{\odot} \geq 12.01$, and the low-mass subsample (blue
points) has $\log M_{\rm\,h}/M_{\odot} < 12.01$.  A 2D
Kolomorov-Smirnov (KS) test yields a $4.2~\sigma$ significance that
the null hypothesis the two subsamples have identical $W_r(2796)$--$D$
distributions is ruled out.  Excluding galaxies with upper limits on
$W_r(2796)$ [absorbers only], the significance is $4.0~\sigma$.
Clearly, there is significant mass segregation in the $W_r(2796)$--$D$
distribution, such that lower mass halos cluster toward smaller
$W_r(2796)$ and $D$, whereas higher mass halos cluster toward larger
values.

Since halo mass is a correlated source of scatter on the
$W_r(2796)$--$D$ plane, we examine $W_r(2796)$ versus $D/R_{\rm vir}$,
since $R_{\rm vir} \propto (M_{\rm\,h})^{1/3}$.  We plot these data in
Figure~\ref{fig:EWvsD}c.  The solid line is the fit $\log W_r(2796) =
\alpha_1 \log (D/R_{\rm vir}) + \alpha_2$, using the Expectation
Maximization-Likelihood method \citep{wolynetz79, isobe86}, which
accounts for upper limits.  The fit is normalized to the mean
$W_r(2796)$ of absorbing galaxies at $D/R_{\rm vir}=0.3$.  We obtained
$\alpha_1 = - 2.04\!\pm\!0.21$ and $\alpha_2 = - 1.60\!\pm\!0.15$.
The dashed lines are $1~\sigma$ uncertainty curves.

A BHK-$\tau$ non-parametric rank correlation test \citep{bhk74,
isobe86}, which accounts for upper limits, yielded an $8.2~\sigma$
significance for the anti-correlation between $W_r(2796)$ and $D$
\citep{nielsen12}.  Using the BHK-$\tau$ test, we find that
$W_r(2796)$ is anti-correlated with $D/R_{\rm vir}$ at a $8.9~\sigma$
significance.  Relative to the fits, the variance of the data on the
$W_r(2796)$--$D$ plane is reduced from 0.50 to 0.11 on the
$W_r(2796)$--$D/R_{\rm vir}$ plane (absorbers only).  An $F$-test
comparing the individual variances, $(\log W_{\rm fit} - \log
W_{i})^2$, in both planes yielded probability $P(F\,)=1\times
10^{-10}$ that their distributions are drawn from the same parent
population; the scatter is {\em significantly\/} reduced in the
$W_r(2796)$--$D/R_{\rm vir}$ plane.  Furthermore, the distribution of
halo masses about the fit has been homogenized.  A 2D KS test
comparing the distributions of the low- and high-mass subsamples is
consistent with their being drawn from the same parent distribution
($1.3~\sigma$).

In summary, there is significant systematic mass segregation on the
$W_r(2796)$--$D$ plane.  However, on the $W_r(2796)$--$D/R_{\rm vir}$
plane, the scatter about the fit is reduced with very high
significance; the anti-correlation is highly significant and the
segregation by mass vanishes. The data suggest that the {\MgII}
absorbing CGM is strongly linked to halo mass such that $W_r(2796)
\propto (D/R_{\rm vir})^{-2}$.  Interestingly, the CGM exhibits
substantial ``patchiness'' for $D/R_{\rm vir} \geq 0.2$, where the
relative number of sight lines with non detections begins to increase.
To investigate if this may be connected to galaxy properties, we
performed KS tests to compared the colors and luminosities of
absorbing galaxies and non-absorbing galaxies with $W_r(2796) <
0.1$~{\AA} and $D/R_{\rm vir} \geq 0.2$; we find no statistical
differences.

\begin{figure*}[thb]
\epsscale{0.7} 
\plotone{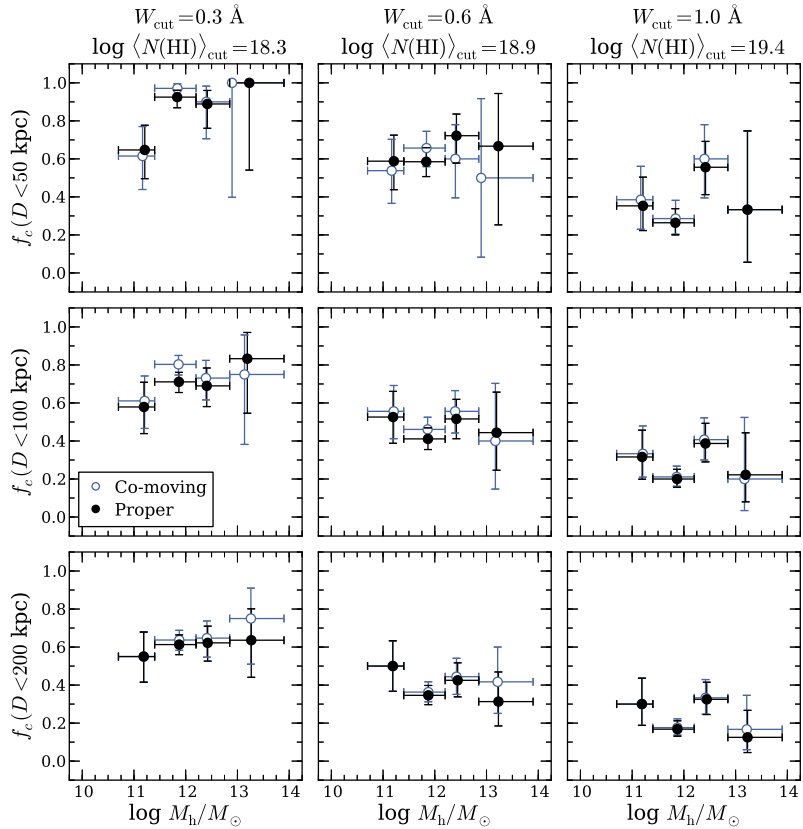}
\caption{The covering fraction, {\fdmax} versus $\log M_{\rm
\,h}/M_{\odot}$ for $W_r(2796) \geq W_{\rm cut}$ inside impact
parameter $D$.  Each $W_{\rm cut}$ corresponds to a mean minimum $\log
N({\HI})$.  Both proper (black points) and co-moving (open blue
points) coordinates are shown.  (upper) $D \leq 50$~kpc. (middle) $D
\leq 100$~kpc.  (lower) $D \leq 200$~kpc.  Contrary to theoretical
expectations for cold-mode accretion, we find that {\fdmax} does {\it
not\/} vanish for $M_{\rm\,h} \geq M_{\rm crit}$.}
\label{fig:fDmax-Mall}
\end{figure*}

\subsection{Covering Fraction versus $D$ and $D/R_{\rm vir}$}

Further insight is provided by the behavior of the covering fraction
of the {\MgII} absorbing CGM, which has been investigated in some
detail \citep{kacprzak08, chen10a, nielsen12}.  \citet{nielsen12}
showed a dependence on galaxy $B$-band luminosity in that higher
luminosity galaxies have higher covering fractions at a given impact
parameter than lower luminosity galaxies.  Since luminosity is a proxy
for mass, their result motivates an examination of the covering
fractions as a function of $M_{\rm\,h}$.

In the upper panels of Figure~\ref{fig:fDbar}, we present {\fdbar},
the CGM covering fraction in an impact parameter bin having $W_r(2796)
\geq W_{\rm cut}$.  The uncertainties for all covering fractions
presented in this work were computed using binomial statistics.  We
computed {\fdbar} for the high-mass (red triangles) and low-mass
(green squares) subsamples and for the full sample (dashed bars).
Generally, high-mass halos have larger {\fdbar} at $D> 50$~kpc with
increasing significance as $W_{\rm cut}$ is lowered.  Note that, for
$D>100$~kpc, low-mass halos have ${\fdbar} = 0$ for all $W_{\rm cut}$,
whereas high-mass halos have ${\fdbar} \simeq 0.3$--0.1.

In the lower panels of Figure~\ref{fig:fDbar}, we plot {\fdrbar}, the
CGM covering fraction in a $D/R_{\rm vir}$ bin with $W_r(2796) \geq
W_{\rm cut}$.  We find that {\fdrbar} for high- and low-mass halos are
statistically identical; {\fdrbar} has no mass dependence for all
$W_{\rm cut}$ at all $D/R_{\rm vir}$.

Using the median fit between $W_r(2796)$ and $N({\HI})$ obtained by
\citet{menard09b}, each $W_{\rm cut}$ converts to a mean minimum
neutral hydrogen column density.  However, we caution that we
extrapolated below the minimum $W_r(2796)$ of their fit ($0.5$~{\AA}).

\begin{figure*}[thb]
\epsscale{0.7} 
\plotone{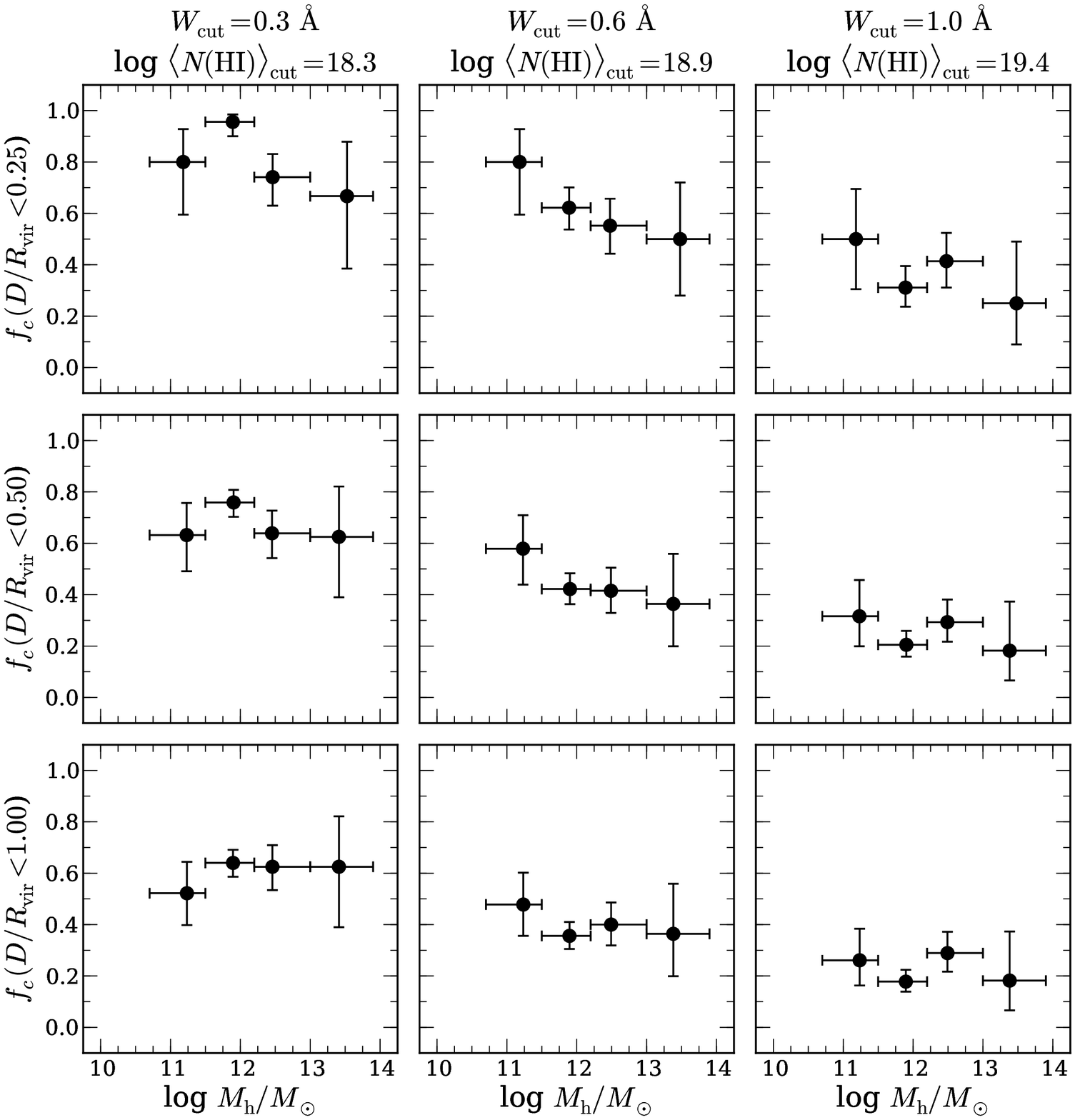}
\caption{The CGM covering fraction, {\fdrmax}, versus $\log M_{\rm
\,h}/M_{\odot}$ for $W_r(2796) \geq W_{\rm cut}$ inside fractional
projection $D/R_{\rm vir}$.  Each $W_{\rm cut}$ corresponds to a mean
minimum $\log N({\HI})$.  (top) $D/R_{\rm vir} \leq 0.25$. (middle)
$D/R_{\rm vir} \leq 0.5$.  (bottom) $D/R_{\rm vir} \leq 1.0$.  Though
the data hint that {\fdrmax} may decrease with increasing $\log M_{\rm
\,h}$, the trends are not statistically significant for our sample.
The data suggest self-similarity of the covering fraction with halo
mass, with {\fdrmax} depending primarily on $W_{\rm cut}$ and the
$D/R_{\rm vir}$ cut.}
\label{fig:fDmaxMRv}
\end{figure*}

The combined behavior of {\fdbar} and {\fdrbar} strongly suggests that
the {\MgII} absorbing CGM is self-similar with halo mass over the
range $10.7 \leq \log M_{\rm\,h}/M_{\odot} \leq 13.9$.  If so, in a
fixed $D$ bin, the higher {\fdbar} values for higher mass halos is
naturally explained by the fact that the CGM of higher mass halos is
probed at smaller $D/R_{\rm vir}$, whereas lower mass halos are probed
at larger $D/R_{\rm vir}$.  



\subsection{Covering Fraction versus $M_{\rm\,h}$}

In Figure~\ref{fig:fDmax-Mall}, we present {\fdmax} versus $M_{\rm
\,h}$, where {\fdmax} is the CGM covering fraction within impact
parameter $D$ with $W_r(2796) \geq W_{\rm cut}$.  We computed {\fdmax}
for both proper coordinates (black points) and for co-moving
coordinates (open blue points).  The behavior of {\fdmax} is not
sensitive to the choice of coordinates.  We find that {\fdmax} shows
no definitive trend with $M_{\rm\,h}$, being consistent with a
constant value within errors.  


We also computed {\fdrmax}, the CGM covering fraction within $
D/R_{\rm vir}$ with $W_r(2796) \geq W_{\rm cut}$.  We plot {\fdrmax}
versus $M_{\rm\,h}$ in Figure~\ref{fig:fDmaxMRv}.  Again, a suggestion
of self-similarity of the {\MgII} absorbing CGM is apparent; {\fdrmax}
is effectively constant with halo mass within errors, primarily
depending on $W_{\rm cut}$ and the $D/R_{\rm vir}$ cut.  Though a
suggestion decreasing {\fdrmax} with increasing $M_{\rm\,h}$ is
visually apparent, non-parametric rank correlation tests indicate that
no correlation with $M_{\rm\,h}$ is consistent with the data (null
hypothesis satisfied within $0.2$--$0.9~\sigma$).

Most importantly, we point out that neither {\fdmax} nor {\fdrmax}
rapidly decline or vanish for $M_{\rm\,h} > M_{\rm crit}$.  This
behavior is in conflict with theoretical predictions in which
cold-mode accretion diminishes for $M_{\rm\,h} > M_{\rm crit}$
\citep[cf.,][]{birnboim03, keres05, dekel06, keres09, stewart11,
vandevoort11}, either implying alternative origins for a ubiquitous
{\MgII} absorbing CGM or that cold mode accretion persists above
$M_{\rm crit}$.

\section{Discussion and Conclusions}
\label{sec:conclude}

We applied the halo abundance matching technique to determine the
galaxy virial masses, $M_{\rm\,h}$, for 183 ``isolated'' galaxies
from {\magiicat} \citep[][]{nielsen13}.  We report four main results:

1. To a significant degree, the substantial scatter about the
$W_r(2796)$--$D$ anti-correlation is explained by a systematic
segregation of halo mass on the $W_r(2796)$--$D$ plane. With
$4.0~\sigma$ significance, the high-mass halos exhibit larger
$W_r(2796)$ at greater $D$ compared to the low-mass halos.  The
significance of the $W_r(2796)$--$D/R_{\rm vir}$ anti-correlation is
$8.9~\sigma$ and the data indicate $W_r(2796) \propto (D/R_{\rm
  vir})^{-2}$.  On the $W_r(2796)$--$D/R_{\rm vir}$ plane, systematic
halo mass segregation vanishes and the scatter is reduced with very
high significance.  These results suggest that {\MgII} absorption
strength is strongly linked to $D/R_{\rm vir}$ via halo mass.

2.  For all $W_{\rm cut}$ and $D/R_{\rm vir}$, there is no dependence
of the covering fraction {\fdrbar} on $M_{\rm\,h}$, whereas for
smaller $W_{\rm cut}$ and $D>50$~kpc, {\fdbar} is higher for the
high-mass halos. Since {\fdrbar} decreases with increasing $D/R_{\rm
  vir}$, the $M_{\rm\,h}$ dependence of {\fdbar} can be explained by
the fact that at a fixed $D$, higher mass halos are probed at smaller
$D/R_{\rm vir}$ and lower mass halos are probed at larger $D/R_{\rm
  vir}$.  The combined behavior of $W_r(2796)$ versus $D/R_{\rm vir}$,
{\fdbar}, and {\fdrbar} strongly suggests that the {\MgII} absorbing
CGM is self-similar with halo mass over the range $10.7 \leq \log
M_{\rm \,h}/M_{\odot} \leq 13.9$.

3.  The covering fraction {\fdmax} is constant with $M_{\rm\,h}$.
Similarly, the covering fraction {\fdrmax} is a constant with
$M_{\rm\,h}$ for all $W_{\rm cut}$ and $D/R_{\rm vir}$ cuts.  The
magnitude of the covering fraction depends only on $W_{\rm cut}$ and
the $D$ and $D/R_{\rm vir}$ cuts, decreasing with increasing $W_{\rm
  cut}$ for fixed $D$ cut and with increasing $D$ cut for fixed
$W_{\rm cut}$.  The lack of dependence of {\fdrmax} with $M_{\rm\,h}$
further supports our conclusion of the self-similarity of the {\MgII}
absorbing CGM with halo mass.

4. Neither {\fdmax} nor {\fdrmax} precipitously drop for $M_{\rm\,h}
\geq M_{\rm crit}$.  This behavior is in direct conflict with
theoretical expectations of cold-mode accretion.  If the {\MgII}
absorbing CGM is self-similar with halo mass, it would imply that
{\HI} in cold gas is also self-similar with halo mass.

If cold-mode accretion substantially diminishes above some $M_{\rm
crit}$ \citep{birnboim03, keres05, dekel06, keres09, stewart11,
vandevoort11}, our results imply that the {\MgII} absorbing CGM must
be sustained by other mechanisms in $M_{\rm\,h} > M_{\rm crit}$ halos.
A most obvious candidate is outflows, which are observed in starburst
galaxies \citep[e.g.,][]{tremonti07, martin09, weiner09, rubin10,
martin12}.  On average, this would imply that, in high-mass halos, the
dynamical and cooling times of non-escaping wind material are balanced
with the wind cycling time scale such that their cold gas reservoirs
are more-or-less observationally indistinguishable from the cold
accretion reservoirs of low-mass halos.  Simulations indicate
recycling time scales on the order of 0.5--1.0 Gyr that decrease with
increasing $M_{\rm\,h}$ \citep{oppenheimer10, stewart11}.  Thus,
galaxies in higher mass halos with quiescent star formation for longer
than $10^9$ yr would not be expected to have detectable {\MgII} or
strong {\HI} absorption \citep[c.f.,][]{churchill-q1317}.

On the other hand, since the number of sub-halos increases in
proportion to halo mass \citep{klypin11}, it is plausible that
absorption from sub-halos counteracts our ability to observe the cut
off of cold-mode accretion in the central halos with $M_{\rm\,h} \geq
M_{\rm crit}$.  Alternatively, the theoretical prediction that
cold-mode accretion universally diminishes in higher mass halos may
not be correct.

Since $R_{\rm vir} \propto (M_{\rm\,h}/\Delta_c)^{1/3}$ and
$W_r(2796)\propto (D/R_{\rm vir})^{-2}$, we deduce that $W_r(2796)
\propto D^{-2}(M_{\rm\,h}/\Delta_c)^{2/3}$ over the range $10.7 \leq
M_{\rm\,h} \leq 13.9$.  That is, $W_r(2796) \propto D^{-2}$ for fixed
$M_{\rm\,h}$ and $W_r(2796) \propto (M_{\rm\,h}/\Delta_c)^{2/3}$ for
fixed $D$.  This behavior is consistent with the halo mass segregation
we observe on the $W_r(2796)$--$D$ plane, and is corroborated by the
$W_r(2796)$-stellar mass correlation reported by \citet{martin12} for
starburst galaxies and the stellar mass dependence on the halo cross
section of {\MgII} absorbing gas \citep{chen10b}.  However, it is
contrary to the global $W_r(2796)$--$M_{\rm\,h}$ anti-correlation
reported by \citet{bouche06} and \citet{gauthier09} based upon
cross-correlation techniques.  We explore implications of this
discrepancy in \citet{churchill13}.

\acknowledgments 

We thank Peter Behroozi for insightful discussions during his visit to
NMSU and Jean-Ren\'{e} Gauthier for valuable input.  CWC and NMN were
partially supported through grant HST-GO-12252 provided by NASA's
Space Telescope Science Institute, which is operated by AURA under
NASA contract NAS 5-26555. NMN was also partially supported through a
NMSGC Graduate Fellowship and a three-year Graduate Research
Enhancement Grant (GREG) sponsored by the Office of the Vice President
for Research at New Mexico State University.  Dedicated to the memory
of William Arthur Churchill.


\begin{thebibliography}{}


\bibitem[Behroozi, Conroy, \& Wechsler(2010)Behroozi{\etal}]{behroozi10} 
Behroozi, P.~S., Conroy, C., \& Wechsler, R.~H.\ 2010, \apj, 717, 379

\bibitem[Bouch{\'e} {\etal}(2006)]{bouche06} Bouch{\'e}, N., Murphy,
M.~T., P{\'e}roux, C., Csabai, I., \& Wild, V.\ 2006, \mnras, 371, 495

\bibitem[Bouch{\'e} {\etal}(2011)]{bouche11} Bouch{\'e}, N., Hohensee,
  W., Vargas, R., {\etal} 2011, MNRAS, 426, 801

\bibitem[Birnboim \& Dekel(2003)]{birnboim03} Birnboim, Y., \& Dekel,
  A.\ 2003, MNRAS, 345, 349

\bibitem[Bolzonella {\etal}(2000)]{bolzonella00} Bolzonella, M.,
  Miralles, J.-M., \& Pell{\'o}, R.\ 2000, \aap, 363, 476

\bibitem[Brown, Hollander, \& Korwar(1974)]{bhk74} Brown, B. W.,
Hollander, M., \& Korwar, R. M. 1974, in Reliability and Biometry, 327

\bibitem[Bryan \& Norman(1998)]{bryan98} 
Bryan, G.~L., \& Norman, M.~L.\ 1998, \apj, 495, 80

\bibitem[Chen {\etal}(2010a)]{chen10a} 
Chen, H.-W., Helsby, J.~E., Gauthier, J.-R., Shectman, S.~A.,
Thompson, I.~B., \& Tinker, J.~L.\ 2010a, \apj, 714, 1521

\bibitem[Chen {\etal}(2010b)]{chen10b} 
Chen, H.-W., Wild, V., Tinker, J.~L., {\etal} 2010b, \apjl, 724, L176

\bibitem[Churchill {\etal}(2012a)]{churchill12}
Churchill, C.~W., Kacprzak, G.~G., Nielsen, N. M., Steidel, C.~C., \&
Murphy, M.~T.\ 2012a, \apj, submitted

\bibitem[Churchill, Kacprzak, \& Steidel(2005)]{cwc-china}
Churchill, C. W., Kacprzak, G. G., \& Steidel, C. C. 2005, in {\it
Probing Galaxies through Quasar Absorption Lines}, IAU 199
Proceedings, eds.\ P. R. Williams, C.--G. Shu, \& B. M\'{e}nard
(Cambridge: Cambridge University Press), p.\ 24

\bibitem[Churchill {\etal}(2012b)]{churchill-q1317} Churchill, C.~W.,
Kacprzak, G.~G., Steidel, C.~C., {\etal.} 2012b, arXiv: 1205.0595

\bibitem[Churchill {\etal}(2013)]{churchill13} Churchill, C. W.,
Nielsen, N. M., Kacprzak, G. G., \& Trujillo-Gomes, S. 2013, ApJ, in
preparation

\bibitem[Churchill {\etal}(2000)]{archiveI}
Churchill, C. W., Mellon, R. R., Charlton, J. C., Jannuzi, B. T.,
Kirhakos, S., Steidel, C. C., \& Schneider, D. P. 2000, ApJS, 130, 91

\bibitem[Churchill {\etal}(1999)]{weakI} Churchill, C. W., Rigby,
J. R., Charlton, J. C., \& Vogt, S. S. 1999, ApJS, 120, 51

\bibitem[Coleman {\etal}(1980)]{cww80} Coleman, G.~D., Wu, C.-C.,
  \& Weedman, D.~W.\ 1980, \apjs, 43, 393

\bibitem[Conroy {\etal}(2006)]{conroy06} Conroy, C., Wechsler, R.~H.,
\& Kravtsov, A.~V.\ 2006, \apj, 647, 201

\bibitem[Conroy \& Wechsler(2009)]{conroy09} Conroy, C., \&
Wechsler, R.~H.\ 2009, \apj, 696, 620

\bibitem[Dekel \& Birnboim(2006)]{dekel06} 
Dekel, A., \& Birnboim, Y.\ 2006, \mnras, 368, 2




\bibitem[Gauthier {\etal}(2009)]{gauthier09} Gauthier, J.-R., Chen, 
H.-W., \& Tinker, J.~L.\ 2009, \apj, 702, 50 




\bibitem[Isobe, Feigelson, \& Nelson(1986)]{isobe86} Isobe, T.,
Feigelson, E. D., \& Nelson, P. I. 1986, ApJ, 306, 490


\bibitem[Kacprzak {\etal}(2008)]{kacprzak08} 
Kacprzak, G.~G., Churchill, C.~W., Steidel, C.~C., \& Murphy, M.~T.\
2008, AJ, 135, 922

\bibitem[Kacprzak {\etal}(2011)]{kcems} Kacprzak, G. G., Churchill,
C. W., Evans, J. L., Murphy, M. T., \& Steidel, C. C. 2011, MNRAS,
416, 3118

\bibitem[Kacprzak {\etal}(2007)]{ggk-morphology} Kacprzak, G.~G.,
Churchill, C.~W., Steidel, C.~C., Murphy, M.~T., \& Evans, J.~L.\
2007, \apj, 662, 909

\bibitem[Kacprzak {\etal}(2012)]{ggk-q1317} Kacprzak, G. G.,
Churchill, C. W., Steidel, C. C., Spitler, L. R, Holtzman, J. A., \&
Bouch\'e, N. A. 2012, MNRAS, in press (arXiv:1208.4098)

\bibitem[Kere{\v s} {\etal}(2009)]{keres09} 
Kere{\v s}, D., Katz, N., Fardal, M., Dav{\'e}, R., \& Weinberg,
D.~H.\ 2009, \mnras, 395, 160

\bibitem[Kere{\v s} {\etal}(2005)]{keres05} 
Kere{\v s}, D., Katz, N., Weinberg, D.~H., \& Dav{\'e}, R.\ 2005,
\mnras, 363, 2

\bibitem[Kim, Goobar, \& Perlmutter(1996)]{kim96}
Kim, A., Goobar, A., \& Perlmutter, S. 1996, PASP, 108, 190

\bibitem[Klypin {\etal}(2011)]{klypin11} Klypin, A.~A.,
Trujillo-Gomez, S., \& Primack, J.\ 2011, \apj, 740, 102



\bibitem[Martin \& Bouch{\'e}(2009)]{martin09} 
Martin, C.~L., \& Bouch{\'e}, N.\ 2009, \apj, 703, 1394 

\bibitem[Martin {\etal}(2012)]{martin12} Martin, C.~L., Shapley, 
A.~E., Coil, A.~L., et al.\ 2012, arXiv:1206.5552 

\bibitem[M{\'e}nard \& Chelouche(2009)]{menard09b} M{\'e}nard, B., \&
Chelouche, D.\ 2009, \mnras, 393, 808


\bibitem[Nielsen, Churchill, \& Kacprzak(2012)]{nielsen12} Nielsen,
N. M., Churchill, C.~W., \& Kacprzak, G.~G. \ 2012, \apj, submitted
(arXiv:1211.1380)

\bibitem[Nielsen, Churchill, \& Kacprzak(2013)]{nielsen13} Nielsen,
N. M., Churchill, C.~W., \& Kacprzak, G.~G. \ 2013, \apjl, in prep

\bibitem[Oppenheimer et al.(2010)]{oppenheimer10} 
Oppenheimer, B.~D., Dav{\'e}, R., Kere{\v s}, D., et al.\ 2010,
\mnras, 406, 2325

\bibitem[Rao \& Turnshek(2000)]{rao00} Rao, S.~M., \&
Turnshek, D.~A.\ 2000, \apjs, 130, 1

\bibitem[Ribaudo {\etal}(2011)]{ribaudo11} Ribaudo, J., Lehner,
N., Howk, J.~C., et al.\ 2011, \apj, 743, 207

\bibitem[Rigby, Charlton, \& Churchill(2002)]{weakII}
Rigby, J. R., Charlton, J. C., \& Churchill, C. W. 2002, ApJ, 565, 743

\bibitem[Rodriguez-Puebla {\etal}(2012)]{rodriguez-puebla12} 
Rodriguez-Puebla, A., Drory, N., \& Avila-Reese, V.\ 2012, arXiv:1204.0804 

\bibitem[Rubin {\etal}(2010)]{rubin10}
 Rubin, K.~H.~R., Weiner, B.~J., Koo, D.~C., {\etal} 2010, \apj, 719, 1503

\bibitem[Stewart {\etal}(2011)]{stewart11} 
Stewart, K.~R., Kaufmann, T., Bullock, J.~S., et al.\ 2011, \apjl,
735, L1


\bibitem[Trujillo-Gomez {\etal}(2011)]{trujillo-gomez11}
Trujillo-Gomez, S., Klypin, A., Primack, J., \& Romanowsky, A.~J.\
2011, \apj, 742, 16

\bibitem[Thom {\etal}(2011)]{thom11} Thom, C., Werk, J.~K., Tumlinson,
J., {\etal} 2011, \apj, 736, 1

\bibitem[Tremonti {\etal}(2007)]{tremonti07} Tremonti, C.~A.,
Moustakas, J., \& Diamond-Stanic, A.~M.\ 2007, ApJL, 663, L77

\bibitem[Tumlinson {\etal}(2011)]{tumlinson11} Tumlinson, J., Thom,
  C., Werk, J.~K., {\etal} 2011, Science, 334, 948


\bibitem[van de Voort {\etal}(2011)]{vandevoort11} van de Voort, F.,
  Schaye, J., Booth, C.~M., Haas, M.~R., \& Dalla Vecchia, C.\ 2011,
  MNRAS, 414, 2458


\bibitem[Wang \& Wells(2000)]{wang} Wang, W., \& Wells, M.~T. 2000,
 
\bibitem[Weiner {\etal}(2009)]{weiner09} 
Weiner, B.~J., et al.\ 2009, ApJ, 692, 187

\bibitem[Wolf {\etal}(2003)]{wolf03} Wolf, C.,
Meisenheimer, K., Rix, H.-W., et al.\ 2003, \aap, 401, 73

\bibitem[Wolynetz(1979)]{wolynetz79} Wolynetz, M. S. 1979, Journal of
the Royal Statistical Society, 28, 195 

\end{thebibliography}
\end{document}